\documentclass[prl,aps,floatfix,twocolumn]{revtex4}%
\usepackage{epsfig}
\usepackage{amsmath}
\usepackage{amsfonts}
\usepackage{amssymb}
\usepackage{graphicx}%
\providecommand{\U}[1]{\protect\rule{.1in}{.1in}}
\begin{document}
\title{Phase-sensitive tests of the pairing state symmetry in Sr$_{2}$RuO$_{4}$}
\author{Igor \v{Z}uti\'{c} and Igor Mazin }
\affiliation{Center for Computational Materials Science, 
Naval Research Laboratory,
Washington, D.C. 20375}

\begin{abstract}
Exotic superconducting properties of Sr$_{2}$RuO$_{4}$ have
provided strong support for an unconventional pairing symmetry. However, the
extensive efforts over the past decade have not yet unambiguously resolved the
controversy about the pairing symmetry in this material. While recent
phase-sensitive experiments using flux modulation in Josephson junctions
consisting of Sr$_{2}$RuO$_{4}$ and a conventional superconductor have been
interpreted as conclusive evidence for a chiral spin-triplet pairing, we
propose here an alternative interpretation. We show that an overlooked chiral
spin-singlet pairing is also compatible with the observed phase shifts in
Josephson junctions
and propose further experiments which would distinguish it from its
spin-triplet counterpart.

\end{abstract}

\maketitle

\vspace{-0.6cm}
Unambiguous determination of the pairing state symmetry is one of the key
steps towards understanding the pairing mechanism in a continously growing
class of unconventional superconductors~\cite{unconv}. Phase-sensitive
experiments, capable of identifying angular dependence of the superconducting
order parameter, have provided a crucial evidence for a dominant $d$-wave
orbital symmetry in cuprate
superconductors~\cite{Wollman1993:PRL,Wei1998:PRL,Klemm}. However, much less
is known for other unconventional superconductors such as heavy fermions,
charge transfer salts,
and cobaltates. In particular, there is compelling evidence for an
unconventional pairing in Sr$_{2}$RuO$_{4}$%
~\cite{Maeno1994:N,Mackenzie2003:RMP}, with the strong possibility of 
spin-triplet superconductivity which would be a solid-state analog of 
superfluid He$^{3}$~\cite{Agterberg1997:PRL}.

In superconductors with inversion symmetry an order parameter (gap matrix) can
be expressed as $\hat{\Delta}(\mathbf{k})=\Delta_{0}(\mathbf{k}) i \hat
{\sigma}_{y}$ for spin singlet and $\hat{\Delta}(\mathbf{k})=\mathbf{\hat
{\sigma}} \cdot\mathbf{d(k)} i \hat{\sigma}_{y}$ for spin-triplet pairing.
Here $\mathbf{\hat\sigma}$ are the Pauli spin matrices and scalar (vector)
$\Delta_{0}$ $(\mathbf{d})$ has even (odd) parity in the wavevector
\textbf{k}. Often the symmetry of both the orbital and the spin part of 
$\hat{\Delta}(\mathbf{k})$
remains to be identified and the lack of related understanding
comes from the difficulty in performing phase-sensitive experiments.

While numerous previous experiments probed the pairing symmetry of
Sr$_{2}$RuO$_{4}$~\cite{Mackenzie2003:RMP}, in this context, recent
phase-sensitive experiments~\cite{Nelson2004:S} that provide 
angle-resolved
information are particularly important. The corresponding measurements were
performed in a superconducting quantum interference device (SQUID) geometry,
consisting of a pair of Au$_{0.5}$In$_{0.5}$/Sr$_{2}$RuO$_{4}$ Josephson
junctions. Since Au$_{0.5}$In$_{0.5}$ is a conventional $s$-wave
superconductor, the observed modulation of critical current in an applied
magnetic field was interpreted as conclusive support for the phase shifts
from an odd-parity spin-triplet pairing in Sr$_{2}$RuO$_{4}$%
~\cite{Nelson2004:S,Rice2004:S}.

A similar SQUID geometry was initially proposed~\cite{Geshkenbein1987:PRB} to
study possible $p$-wave pairing in heavy fermions and later also used for
identifying $d$-wave pairing in cuprates~\cite{Wollman1993:PRL}. The critical
current $I_{c}$ is modulated in the applied magnetic field as~\cite{typo}
\begin{equation}
I_{c}\propto\cos(\Phi/\Phi_{0}+\delta_{12}/2), \label{eq:squid}%
\end{equation}
where $\Phi$ is the flux threading the SQUID, $\Phi_{0}$ is the flux quantum,
and $\delta_{12}$ is the intrinsic phase shift of the order parameter between
the two tunneling directions. For a conventional $s$-wave SQUID $\delta
_{12}=0$ and $I_{c}$ has a maximum at $\Phi=0$. In contrast, a phase shift
$\delta_{12}=\pi$, characteristic of unconventional pairing~\cite{pi},
yields a minimum $I_{c}$ at $\Phi=0$. The modulation of external flux together
with the fabrication of junctions with varying tunneling directions in SQUID
geometry therefore provides an angle-resolved phase-sensitive information about
the superconducting pairing symmetry.

The suggested chiral $p$-wave (C\textit{p}W) state with the triplet
order parameter~\cite{Ishida1998:N},
\begin{equation}
\mathbf{d(k)}\propto(k_{x}+ik_{y})\mathbf{\hat{z}}, \label{eq:CpW}%
\end{equation}
in which the spins of the Cooper pairs lie in the RuO$_2$ plane
($\perp {\bf d}$),
is indeed compatible with the experiment~\cite{Nelson2004:S}. However, 
we show here
that it is not the only candidate. There exists another pairing state, allowed
by the tetragonal symmetry of Sr$_{2}$RuO$_{4}$, the singlet chiral $d$-wave
(C\textit{d}W) state $^{1}E_{g}(c)$ with $\Delta_{0}(\mathbf{k})\propto
(k_{x}+ik_{y})k_{z}$, or, more accurately~\cite{Annet1990:AP}
\begin{equation}
\Delta_{0}(\mathbf{k})\propto(k_{x}+ik_{y})\sin k_{z}c, \label{eq:CdW}%
\end{equation}
which is equally consistent~\cite{T-note} with the phase shifts observed in
Ref.~\cite{Nelson2004:S}. We use our findings to propose an experimental test
which would discriminate between C\textit{p}W and C\textit{d}W pairing symmetries.

Could experimental and theoretical reasons be used to rule out C\textit{d}W
and favor only the C\textit{p}W state? The two main arguments in favor of the
C\textit{p}W come from 
muon spin resonance
and Knight shift
experiments~\cite{Mackenzie2003:RMP,Luke1998:N}. The former indicate a
time-reversal symmetry breaking below the transition temperature $T_{c}$, 
incompatible with the 
$d_{x^2-y^2}$-wave state 
in cuprates, 
but fully compatible with either
C\textit{p}W or C\textit{d}W symmetry. The Knight shift ($K$) was initially
interpreted as firm evidence for a triplet state with in-plane spins
(like C\textit{p}W), since no change of the in-plane spin susceptibility below
$T_{c}$ was found. 

Even in singlet superconductors 
($e.g.$, vanadium),
$K$ could remain invariant below $T_{c}$. 
Such behavior is usually attributed to (a)
spin-orbit induced spin-flip scattering which
suppresses the effect of the superconductivity on $K$ and (b) 
an accidental cancellation of the spin, dipole, and orbital
contributions of the Fermi-level electrons to $K$ which leaves only
superconductivity-insensitive contributions such as the Van Vleck
susceptibility. However, a quantitative analysis~\cite{Pavarini2005:P} shows
that the spin-orbit coupling in Sr$_{2}$RuO$_{4}$ is too weak for 
scenario (a) while the accidential cancellation, required for scenario (b), 
does not occur~\cite{Knight}.
Thus, neither of the two explanations of a constant $K$ arising from
singlet pairing is applicable. 

This would have made the Knight shift argument for C{\it p}W very 
convincing, if not for the recent
experiment showing the same result in a field perpendicular to the
plane~\cite{Murukawa2004:PRL}. It was proposed~\cite{Murukawa2004:PRL} 
that the testing field of
0.02 T may be enough to induce a phase transition from the C\textit{p}W 
in Eq.~(\ref{eq:CpW})
to a state with
\textbf{d}$\Vert\mathbf{\hat{x}}$. However, this is highly unlikely: (i) 
the \textbf{d}$\Vert\mathbf{\hat{x}}$ state would have an additional 
horizontal line node, as compared to the
\textbf{d}$\propto(k_{x}+ik_{y})\mathbf{\hat{z}}$ state and therefore lose 
a large part of the pairing energy ( $\sim$ $\Delta$ per electron,
$\Delta\gtrsim1.4$ K $\gg\mu_{B}\times$0.02 T);
(ii) although in the \textbf{d}$\Vert\mathbf{\hat{x}}$ state
the spins of the pairs lie in the $yz$ plane, there is no $y-z$ symmetry (as
opposed to the $xy$ plane) and it is not $a$ $priori$ clear whether the
magnetic susceptibility of the Cooper pair will be the same as for the normal
electrons. Since \textbf{d}$\Vert\mathbf{\hat{x}}$ is
not allowed for a tetragonal symmetry,
it may only appear as a result of a second phase transition below
$T_{c}$, which has never been observed in Sr$_{2}$RuO$_{4}$; (iii) the 
spin-orbit part of the pairing interaction, which keeps the spins in the $xy$ 
plane, despite $z$ being the easy magnetization axis~\cite{Maeno1997:JPSJ},
would have to be weaker than $\mu_{B}\times$0.02 T=1.1 $\mu$eV=0.013 K, an
energy scale much too small for the spin-orbit coupling in Sr$_{2}$RuO$_{4}$.
So, neither the old theories for the lack of a Knight shift reduction below
$T_{c}$, nor the new explanation in terms of the magnetic-field rotated order
parameter withstand quantitative scrutiny; 
the Knight shift in Sr$_{2}$RuO$_{4}$
remains a challenge for theorists. Until this puzzle is resolved, we cannot
use the Knight shift argument for the pairing symmetry determination.

\begin{figure}[ptb]
\caption{ Schematic sample and the Fermi surface geometry for phase-sensitive
SQUID measurements from Ref.~\cite{Nelson2004:S}. (a) Au$_{0.5}$In$_{0.5}%
$/Sr$_{2}$RuO$_{4}$ junction geometry with an interface normal $\mathbf{n}$.
(b) Possible deviation of $\mathbf{n}$ from the $ab$ crystallographic plane.
(c) Warping of the Sr$_{2}$RuO$_{4}$ Fermi surface. The magnitude of the Fermi
wavevector $\mathbf{k}_{F}$ is generally different from the one corresponding
to the cylindrical Fermi surface ($k_{F0})$. }%
\label{fig:1}
\centerline{\psfig{file=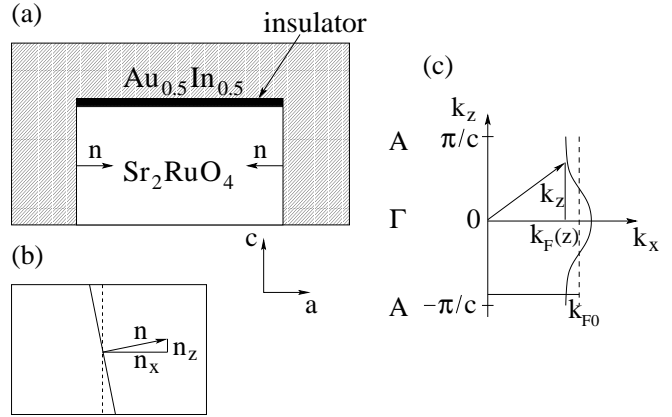,width=1\linewidth,angle=0}}\end{figure}

We now turn to the experiments of Ref.~\cite{Nelson2004:S} and compare
Josephson tunneling between an $s$-wave superconductor and either (a) an even
parity (spin singlet) superconductor or (b) an odd parity (triplet)
superconductor. In the first case, the Josephson current between a
conventional $s$-wave superconductor and an unconventional spin-singlet
superconductor, represented by the order parameters $\Delta_{s-\mathrm{wave}}$
and $\Delta_{0}(\mathbf{k})$, respectively, can be expressed
as~\cite{Geshkenbein1986:PZETF}
\begin{equation}
J \propto\left\langle T_{\mathbf{k}}\mathrm{Im}[\Delta
_{s-\mathrm{wave}}^{\ast}\Delta_{0}(\mathbf{k})]\right\rangle _{FS},
\label{eq:jofk}%
\end{equation}
which depends on the relative phase between the superconducting
order parameters.
The averaging is over all states at the Fermi surface (FS) where the Fermi
velocity, $\mathbf{v}_{F}$, has a positive projection on the tunneling
direction represented by the unit normal \textbf{n} (perpendicular to the
interface plane, see Fig.~\ref{fig:1}) and $T_{\mathbf{k}}$ is the tunneling
probability. For a thick rectangular barrier of width $w$ and height
$U$~\cite{Mazin2001:EPL} we can obtain
\begin{equation}
T_{\mathbf{k}}=\frac{16m^{2}\kappa^{2}v_{L}v_{R}\exp[-2\kappa w-k_{\Vert}%
^{2}w/\kappa]}{(\kappa^{2}+m^{2}v_{L}^{2})(\kappa^{2}+m^{2}v_{R}^{2})},
\label{eq:tkthick}%
\end{equation}
where $\kappa=\sqrt{2m(U-\mu)}$
such that $w\kappa\gg1$ (in the thick barrier limit), $m$ is the free-electron
mass, $\mu$ is the Fermi energy, and we set $\hbar=1$. We use $v_{L,R}$ to
denote normal components of the Fermi velocities in the two superconductors
and $k_{\Vert}$ is the component parallel to the interface. From
Eq.~(\ref{eq:tkthick}) we see that $T_{\mathbf{k}}$ is sharply peaked when
$\mathbf{v}_{F}\parallel\mathbf{n}$.

In the second case, the Josephson current between a singlet and a triplet
superconductor becomes~\cite{Geshkenbein1986:PZETF,Millis1988:PRB}
\begin{equation}
J\propto\left\langle \tilde{T}_{\mathbf{k}}\mathrm{Im}[\Delta
_{0}^{\ast}(\mathbf{k})\mathbf{d(k)}\cdot(\mathbf{n}\times\mathbf{k}%
)]\right\rangle _{FS}, \label{eq:MRS}%
\end{equation}
where we use $\tilde{T}_{\mathbf{k}}$ to denote that, unlike 
$T_{\mathbf{k}}$, it contains matrix elements
corresponding to spin-flip tunneling,
for example, due to magnetic interfaces or spin-orbit coupling. For
nonmagnetic barriers and in the absence of spin-orbit,
there is no spin-flip scattering, therefore
$\tilde{T}_{\mathbf{k}%
}=0$ and the Josephson current vanishes
identically~\cite{Geshkenbein1986:PZETF,Millis1988:PRB,Pals1977:PRB}.

From
Eqs.~(\ref{eq:jofk}) and (\ref{eq:MRS}) we can directly infer that for c-axis
tunneling, $\mathbf{n}\parallel\mathbf{c}$, $J=0$ for both
C\textit{d}W and C\textit{p}W states [$\int dk_{x}dk_{y}(k_{x}+ik_{y})=0$].
For tunneling \textit{precisely} in the $ab$ plane, the current is also zero
for C\textit{d}W while for C\textit{p}W it only vanishes at $\mathbf{n}%
\parallel\mathbf{k}$.

We consider a model of a quasi-two-dimensional (2D) layered superconductor 
which has
a nearly cylindrical FS with a small $c$-axis dispersion originating from the
weak inter-layer hopping. In Fig.~\ref{fig:1}(a) we show the sample geometry
used Ref.~\cite{Nelson2004:S} and in Fig.~\ref{fig:1}(b)
represent a warping of the Fermi surface. In the first approximation for
Sr$_{2}$RuO$_{4}$ such a warping can be expressed as~\cite{Bergemann2003:AP}
\begin{equation}
\mu=\frac{k_{F}^{2}(z)}{2m}[1+\varepsilon\cos k_{z} c], \label{eq:mu}%
\end{equation}
where $|\varepsilon| \ll1$ is the warping parameter ($\varepsilon\approx-7
\times10^{-4}$ \cite{Bergemann2003:AP}), $c$ is the lattice constant along the
crystallographic $z$-direction, and $\mathbf{k}_{F}(z)$ is the $z$-dependent
projection of the Fermi wavevector in the $xy$ plane [see Fig.~\ref{fig:1}%
(c)]. It is convenient to resolve the Fermi wavevector in cylindrical
coordinates ($k_{F}(z)$, $\varphi$, $k_{z}$) with
\begin{equation}
k_{F}(z)=k_{F0}/[1+\varepsilon\cos k_{z} c ]^{1/2}, \label{eq:kF}%
\end{equation}
where $k_{F0}=(2m \mu)^{1/2}$ and for $\varphi=0$ [see Fig.~\ref{fig:1}(c)]
$k_{F}(z)\rightarrow k_{Fx}$.

Schematic representation of the SQUID geometry in Fig.~\ref{fig:1}(a) (adapted
from Ref.~\cite{Nelson2004:S}) is an oversimplification. While efforts were
made to fabricate edges either precisely parallel or perpendicular to the
$c$-axis, in the actual samples the direction of interface planes or their
corresponding normals changes gradually from $a$ to $c$-direction. In several
samples~\cite{Nelson2004:S} an interface nearly parallel to the $ab$ plane, at
Au$_{0.5}$In$_{0.5}$/Sr$_{2}$RuO$_{4}$ junction, was covered by an insulating
oxide [see Fig.~\ref{fig:1}(a)]. It is then plausible to expect that the
normal to such interface could deviate from the $ab$ plane. In
Fig.~\ref{fig:1}(b) we depict a generalized situation in which an interface
normal, $\mathbf{n}=(n_{\rho},n_{\varphi},n_{z})$ with $|n_{z}|\ll1$, need not
lie exactly in the crystallographic $ab$ plane of Sr$_{2}$RuO$_{4}$. We show
below that the analysis of phase-sensitive measurements in terms of the two
small but \emph{finite} parameters, $\varepsilon$ and $n_{z}$, can provide a
qualitatively different interpretation from those which a priori assume
$\varepsilon=n_{z}\equiv0$.

For a conventional superconductor with the FS larger than the one of Sr$_{2}%
$RuO$_{4}$, the Josephson tunneling across a thick rectangular barrier 
can be
obtained from Eqs.~(\ref{eq:jofk}) and (\ref{eq:tkthick}) as 
\begin{align}
J &  \propto\int_{\mathbf{v}_{F}\cdot{\mathbf{n}}>0}%
d\mathbf{k}\delta(\epsilon_{\mathbf{k}}-\mu)\mathbf{v}_{F}\mathbf{\cdot n}%
\exp(-k_{\parallel}^{2}w/2\kappa)\nonumber\\
&  \times\mathrm{Im}(k_{x}+ik_{y})\sin k_{z}c, \label{eq:jk}%
\end{align}
where $k_{\parallel}^{2}=\mathbf{k}_{F}^{2}-(\mathbf{k}_{F}\cdot
\mathbf{n})^{2}$, $\kappa^2 \gg m v_{L,R}^2$, 
and the projection of the Fermi velocity in Sr$_{2}$RuO$_{4}$
along $\mathbf{n}$ is
\begin{equation}
\mathbf{v}_{F}\mathbf{\cdot n}=\frac{k_{F0}[1+\varepsilon\cos k_{z}c]^{1/2}%
}{m}n_{\rho}-\frac{k_{F0}^{2}\varepsilon c\sin k_{z}c}{2m[1+\varepsilon\cos
k_{z}c]}n_{z}. \label{eq:vf}%
\end{equation}
For a thick barrier, the integration can be simplified by noting that the
dominant contribution comes from $k_{\parallel}=0$. The right hand side of
Eq.~(\ref{eq:jk}), in the leading order in $\varepsilon$ and $n_{z}$, 
can be then reduced to 
$\sqrt{\pi\kappa/w}ck_{F0}^{2}n_{z}(1-\varepsilon)$, such that
\begin{equation}
J_{\: \sqcap}=A \: n_{z}(1-\varepsilon), \label{eq:thick}%
\end{equation}
where $A$ characterizes the normal state barrier transparency. Thus, with a
tilted interface ($n_{z}\neq0$) there is a finite Josephson current even in
the absence of any Fermi surface warping ($\varepsilon=0$).

To verify that our findings of finite Josephson current in C\textit{d}W state
are not limited to the specific assumption of a thick rectangular barrier, we
also consider the rather different case of a strong $\delta$-function
barrier. The corresponding transmission probability
is~\cite{Zutic1999:PRB,Mazin2001:EPL}
\begin{equation}
T_{\mathbf{k}}=\frac{4v_{L}v_{R}}{(v_{L}+v_{R})^{2}+4U^{2}},\label{eq:tkdelta}%
\end{equation}
where $v_{L,R}$ are the normal components of the Fermi velocities in the two
superconductors and $U$ ($\gg v_{L,R}^{2}$) is the scattering strength.
From Eqs.~(\ref{eq:jofk}),~(\ref{eq:tkdelta}) we obtain
\begin{equation}
J\propto\int_{\mathbf{v}_{F}\mathbf{\cdot n}>0}d\mathbf{k}%
\delta(\epsilon_{\mathbf{k}}-\mu)\mathbf{v}_{F}\mathbf{\cdot n} \: \mathrm{Im}%
(k_{x}+ik_{y})\sin k_{z}c,\label{eq:jofk3}%
\end{equation}
where, unlike in the case of a thick barrier, we perform
 $\varphi$ and $k_{z}$ integration.
In the leading order, the right hand side of Eq.~(\ref{eq:jofk3}) is $-\pi
k_{F0}^{2}\varepsilon n_{z},$ and yields
\begin{equation}
J_\delta =-A \: \varepsilon n_{z},\label{eq:delta}%
\end{equation}
where again $A$ characterizes the normal state transparency. In contrast to
the thick-barrier result, the current now vanishes in the absence of FS
warping. From Eqs.~(\ref{eq:thick}),~(\ref{eq:delta}) one can conjecture 
that for a
general case $J\approx A(s-\varepsilon)$, where $0\lesssim s\lesssim1$.

The presence of small parameters $\varepsilon$ and $n_{z}$ in
Eqs.~(\ref{eq:thick}),~(\ref{eq:delta}) shows that the Josephson current in the
C\textit{d}W state would be reduced as compared to the conventional SQUID with
$s$-wave electrodes. However, the alternative picture, based on the
C\textit{p}W state, also contains small parameters which should be kept in mind
when interpreting the experiment of Ref.~\cite{Nelson2004:S}. In addition to
the small relative strength of the spin-orbit coupling (quantified by the
admixture of $S_{\downarrow}$ into a nonrelativistic $S_{\uparrow}$ state,
or the spin-orbit induced band shift relative to the band width
\cite{small}), there can also be another small factor --- a ratio of the 
lattice constant and the superconducting coherence length~\cite{Fenton1985:SSC},
approximately $6\times10^{-3}$ \cite{Mackenzie2003:RMP}.

Results from Eq.~(\ref{eq:thick}),~(\ref{eq:delta}) confirm that the 
C\textit{d}W
state could be compatible with the phase shifts observed in
Ref.~\cite{Nelson2004:S}. Furthermore, the azimuthal dependence of an order
parameter coincides for both C\textit{d}W and C\textit{p}W states.
While the proposed symmetry
arguments~\cite{Nelson2004:S,Rice2004:S} exclude most of superconducting
states allowed in the tetragonal symmetry~\cite{Asano2005:P}, 
these arguments alone are not
sufficient to unambiguously identify the odd-pairing of C\textit{p}W state.
Instead, to confirm that a C\textit{p}W state has indeed been observed, one
would need to accurately calculate the expected magnitudes of the Josephson 
current for both chiral states. In particular we propose a modification of
the experimental configuration~\cite{Nelson2004:S} such that the
interface plane would be slanted at $\approx45^{o}$ with the $c$-axis. If the
corresponding ratio of the Josephson current to the normal state conductivity
becomes substantially larger ($n_{z}$ is no longer small) than in 
Ref.~\cite{Nelson2004:S},
it would be  strong support for the chiral singlet state in 
Eq.~(\ref{eq:CdW}).

Another important distinction between the two
chiral states is the presence of nodes in the superconducting gap. 
In contrast to the C{\it p}W state, C{\it d}W requires by symmetry 
a horizontal line node
[see Eqs.~(\ref{eq:CpW}),~(\ref{eq:CdW})]. 
The idea of a horizontal line node~\cite{Mackenzie2003:RMP} has
been entertained by experimentalists~\cite{Tanatar2001:PRL} and
theorists~\cite{Zhitomirsky2001:PRL} 
for a while, although recently it has fallen
out of favor. Still, some researchers insist on the existence of a horizontal
line node~\cite{Contreras2004:PRB}. Moreover, in the Josephson experiments of
Ref.~\cite{Nelson2004:S} evidence was found for a substantial $k_{z}$
dependence, albeit not necessarily for horizontal nodes, 
of the order parameter in
Sr$_{2}$RuO$_{4}$~\cite{Liu2005:PC}. How could such a material with nearly 2D
electronic structure develop a highly 3D superconducting state? To answer
this question we point out the following facts: (a) practically no
ferromagnetic spin fluctuations, favorable for a $p$-wave pairing, have been
experimentally found in Sr$_{2}$RuO$_{4}$; (b) antiferromagnetic spin
fluctuations at \textbf{q}$=(2/3,2/3,q_{z})$ have negligible $z$
dispersion; (c) the crystal structure of Sr$_{2}$RuO$_{4}$, as opposed to its
electronic structure, is fairly 3D, so one can expect the electron-phonon
coupling to be quite 3D as well (d) there is a sizeable O isotope effect in
Sr$_{2}$RuO$_{4}$, which strongly changes with introduction of pair-breaking
symmetries~\cite{Mao2001:PRB}. While electron-phonon coupling $per$ $se$ 
can only induce an $s$-wave pairing, such a pairing would be prevented by
the strong antiferromagnetic spin fluctuations. However, for the 
proposed C\textit{d}W
state, any 2D interaction cancels out, including the magnetic
interaction. Should the electron-phonon coupling have a maximum say, at
$(1/2,1/2,1/2)$, as opposed to $(1/2,1/2,0)$, the C\textit{d}W state would
have been immediately stabilized providing a plausible scenario for
spin-singlet superconductivity in Sr$_{2}$RuO$_{4}$.

In conclusion, we have revealed that a completely overlooked chiral $d$-wave 
pairing state in Sr$_{2}$RuO$_{4}$ 
is equally compatible with the existing body of experimental data as 
the generally accepted chiral $p$-wave state.
We have proposed phase-sensitive experiments in a SQUID geometry
with a variable tilting angle capable of unambiguously distinguishing
between the two chiral states. 

We thank R. A. Klemm, Y. Liu, and D. J. Van Harlingen for useful discussions.
This work was supported by the US ONR. I. \v{Z}. acknowledges financial
support from the National Research Council.


\begin{references}
\bibitem{unconv}
In an unconventional superconductor, often synonymous to
a non $s$-wave orbital symmetry, there is at least one more
symmetry broken in addition to the gauge symmetry.
\end{references}
\end{document}